\documentclass[a4paper,12pt]{article}
\usepackage{amsmath,amsfonts,amssymb}

\begin{document}

\begin{center}
{\Large\bf The masses of vector mesons
in holographic QCD at finite chiral chemical potential}
\end{center}

\begin{center}
{\large S. S. Afonin$^a$, A. A. Andrianov$^a$, D. Espriu$^b$}
\end{center}

\begin{center}
{\small $^a$ V. A. Fock Department of Theoretical Physics,
Saint-Petersburg State University, 1 ul. Ulyanovskaya, 198504 St.
Petersburg, Russia\\
$^b$ Departament d'Estructura i Constituents de la Mat\`{e}ria\\ and
Institut de Ci\`{e}ncies del Cosmos (ICCUB),\\
Universitat de Barcelona, Marti Franqu\`{e}s 1, 08028 Barcelona,
Spain}
\end{center}

\begin{abstract}
Central heavy-ion collisions may induce sizeable
fluctuations of the topological charge. This effect is expected to
distort the dispersion relation for the hadron masses. We
construct a general setup for a compact description of this
phenomenon in the framework of bottom-up holographic approach to
QCD. A couple of soft wall holographic models are proposed for the
vector mesons. The states having different circular polarizations
are shown to have different effective mass. The requirement of
stability imposes strong constraints on the possible choice of
models.
\end{abstract}

\section{Introduction}

Recently the study of light mesons by means of a bottom-up
holographic approach has attracted a lot of attention(see, e.g.,
the short surveys~\cite{rev_hol}). Usually the spectra of resonances
and the related physics are analysed at vacuum conditions.
However, in view of current experiments with relativistic heavy
ion collisions at RHIC, GSI, and CERN, it is useful to include
in the holographic models some non-trivial
external conditions. One of theoretical methods consists in
incorporation of the Chern-Simons (CS) term in a holographic
action. This permits to address holographically such problems as
the magnetic susceptibility of the quark condensate~\cite{Gor1},
the chiral magnetic effect~\cite{Gor2,Rubakov}, some subtle
questions in the behavior of the correlation
functions~\cite{Col1,Iat}, and derivation of the $O(p^6)$ Chiral
Perturbation Theory low-energy constants~\cite{Col2}.

It has been demonstrated recently~\cite{And2} that a
suitable CS term may lead to interesting meson
phenomenology. This CS term is motivated by the possibility
of local parity breaking taking
place in baryonic matter; a phenomenon that could be
triggered by large topological fluctuations taking place
in central heavy ion collisions. For light quarks
in a quasi-equilibrium situation the
creation of a topological charge translates immediately into
the generation of a finite chiral chemical potential

Inspired by these observations, in the present paper we analyse
the impact of the CS term in the background considered in
Ref.~\cite{And2} on the mass spectrum of the vector mesons in the
Soft Wall (SW) holographic model~\cite{son2}. This will give a
generalization of results of Ref.~\cite{And2} to higher radial
excitations in the large-$N_c$ limit of QCD. Our choice of the
model is motivated by its nice property of possessing the linear
Regge like spectrum which is expected in the first approximation
both experimentally~\cite{phen} and in the string like models of
hadrons. To make the consideration clear we will restrict
ourselves by the simplest version of the SW model.

The paper is organized as follows. The general holographic setup
is presented in Section~2. The analysis of impact of the axial
chemical potential on the mass spectrum is given in Section~3. A
couple of exactly solvable models are constructed in Section~4. A
short discussion of our results are contained in the concluding
Section~5.

\section{The holographic setup}

We consider the gauge SW model~\cite{son2} with 5D Abelian fields
$L$ and $R$ dual (on the AdS$_5$ boundary) to the sources of the
left and right 4D vector currents. A parity-odd 5D CS term is added
to the action,
\begin{equation}
\label{01}
S=S_{\text{free}}[L]+S_{\text{free}}[R]+S_{\text{CS}}[L]-S_{\text{CS}}[R],
\end{equation}
\begin{equation}
\label{02}
S_{\text{free}}[B]=-\frac{1}{8g_5^2}\int d^4x dz e^\varphi\sqrt{g}B_{MN}B^{MN},\qquad B=L,R,
\end{equation}
\begin{equation}
\label{03}
S_{\text{CS}}[B]=-k\int d^4x dz\epsilon^{MNABC}B_M B_{NA}B_{BC}.
\end{equation}
The 5D space is the AdS$_5$ one with the metric
\begin{equation}
\label{2}
ds^2=\frac{R^2}{z^2}(dx_{\mu}dx^{\mu}-dz^2),\qquad \mu=0,1,2,3,
\end{equation}
where $R$ is the radius of the AdS$_5$ space and $z>0$ represents
the holographic coordinate. The dilaton background $e^\varphi$ is
not yet fixed for generality. As usual, the constants $g_5$ and
$k$ are fixed by matching to the ultraviolet asymptotics of the
two-point vector correlator~\cite{son1} and to the axial anomaly,
respectively
\begin{equation}
\label{04}
\frac{g_5^2}{R}=\frac{12\pi^2}{N_c},\qquad k=\frac{N_c}{24\pi^2}.
\end{equation}
In terms of the vector, $V=(L+R)/2$, and axial-vector,
$A=(R-L)/2$, fields the free and CS parts of the action can be
rewritten as (Lorentz indices are lowered)
\begin{equation}
\label{05}
S_{\text{free}}=-\frac{1}{4g_5^2}\int d^4x dz \frac{e^\varphi}{z}\left(V_{MN}^2+A_{MN}^2\right),
\end{equation}
\begin{equation}
\label{06}
S_{\text{CS}}=-k\int d^4x dz\epsilon^{MNABC}A_M\left(V_{NA}V_{BC}+A_{NA}A_{BC}\right).
\end{equation}

If one wishes to provide conservation of the 4D vector current,
the Bardeen surface counterterm must be added,
\begin{equation}
\label{07}
S_{\text{B}}=2k\int d^4x \epsilon^{\mu\nu\lambda\rho}A_\mu V_\nu V_{\lambda\rho}.
\end{equation}
Then one obtains the standard result for the covariant anomaly~\cite{Rebhan},
\begin{equation}
\label{08}
\partial_\mu J_\mu^V=0,\qquad \partial_\mu J_\mu^A=3kV_{\mu\nu}\tilde{V}_{\mu\nu}+kA_{\mu\nu}\tilde{A}_{\mu\nu},
\end{equation}
where $\tilde{F}_{\mu\nu}=\frac12\epsilon^{\mu\nu\lambda\rho}F_{\lambda\rho}$.

In contrast to the finite density effects, the thermal corrections
to the meson masses appear in the next-to-leading order in the
large-$N_c$ counting as they emerge due to the pion loops. On the
other hand, the holographic approach is inherently large-$N_c$
one. For this reason we will not include the finite temperature
effects into our considerations.

\section{Embedding the axial chemical potential}

In Ref.~\cite{And2} it was assumed that the axial chemical
potential $\mu_5$ arises from a time-dependent but spatially
homogeneous background of a pseudoscalar field $a(t)$ such that
$\mu_5=\dot{a}(t)$. In the 5D setup, we will treat $A_M$ as a
background axial-vector field and relate $a(t)$ to the
$z$-component of $A_M$. Namely, we assume for the vacuum
expectation value of the vector field that
\begin{equation}
\label{09}
\langle A_M\rangle=\langle A_z\rangle=\mu_5x_0 f(z),
\end{equation}
where the shape function $f(z)$ will be specified later. The
expressions~\eqref{08} and~\eqref{09} lead to the following form
for the vector part of the action~\eqref{01},
\begin{equation}
\label{1}
S=\frac{R}{g_5^2}\int d^4x dz\left(-\frac{e^\varphi}{4z}V_{MN}^2+
\xi\mu_5f\epsilon^{05ABC}V_A\partial_BV_C\right).
\end{equation}
Here $\xi=\frac{2kg_5^2}{R}=1$ (see~\eqref{04}).

In the axial gauge $V_z=0$, the equation of motion reads
\begin{equation}
\label{4a}
\partial_z\left(\frac{e^{\varphi}}{z}\partial_z V_\mu\right)-\frac{e^{\varphi}}{z}\partial_\mu^2V_\mu-
2\mu_5f\epsilon_{mik}\partial_iV_k=0.
\end{equation}
The small latin indices denote the usual space coordinates,
$m,i,k=1,2,3$. Making the 4D Fourier transform, $V_\mu(x,z)=\int
d^4p e^{ipx}V_\mu(p,z)$ and assuming the standard plane wave
ansatz, $V_\mu(p,z)=\varepsilon_\mu v(z)$, we arrive at the
equation for the particle-like excitations,
\begin{equation}
\label{4}
\left[\partial_z\left(\frac{e^{\varphi}}{z}\partial_z
v\right)+\frac{e^{\varphi}}{z}p^2v\right]\varepsilon_{\mu}+
i2\mu_5f\epsilon_{mik}p_i\varepsilon_k v=0.
\end{equation}
The physical spectrum is given by the eigenvalues $p_n^2=m_n^2$ of normalizable
solutions of Eq.~\eqref{4}. However, the last term
induces the mixing between different polarizations. We must find a
basis diagonalizing the Eq.~\eqref{4}.

For convenience, let us introduce the notation for the differential
operator
\begin{equation}
\label{5}
\hat{F}=\partial_z\left(\frac{e^{\varphi}}{z}\partial_z
\right)+\frac{e^{\varphi}}{z}p^2.
\end{equation}
The space-like part of Eq.~\eqref{4} can be rewritten in the vector form
\begin{equation}
\label{6}
\left(\hat{F}\vec{\varepsilon}+i2\mu_5f\vec{p}\times\vec{\varepsilon}\right)v=0.
\end{equation}
The equation~\eqref{6} is diagonalized with the help of the
projectors,
\begin{equation}
\label{7}
\hat{P}^{\parallel}_{ik}=\frac{p_ip_k}{\vec{p}^2},
\end{equation}
\begin{equation}
\label{8}
\hat{P}^{\pm}_{ik}=\frac12\left[\delta_{ik}-\frac{p_ip_k}{\vec{p}^2}\pm\frac{i}{|\vec{p}|}\epsilon_{ikn}p_n\right].
\end{equation}
The projectors~\eqref{8} on the "circular" polarizations have the following evident properties,
\begin{equation}
\label{9}
\hat{P}^{\pm}\hat{P}^{\mp}=0,\quad \hat{P}^{\pm}\hat{P}^{\pm}=\hat{P}^{\pm},\quad \text{tr}\hat{P}^{\pm}=1,\quad
\hat{P}^{\pm}\vec{p}=0,\quad \hat{P}^{+}+\hat{P}^{-}=\hat{P}^{\perp}.
\end{equation}
Now we change the basis for the space-like polarizations from
$\vec{\varepsilon}=(\varepsilon_1,\varepsilon_2,\varepsilon_3)$ to
$(\varepsilon_{\parallel},\varepsilon_{-},\varepsilon_{+})$. The
equation~\eqref{6} takes the form
\begin{equation}
\label{10}
\hat{F}v^{\parallel}=0,
\end{equation}
\begin{equation}
\label{11} \left(\hat{F}\pm2\mu_5f|\vec{p}|\right)v^{\pm}=0.
\end{equation}
Thus, the longitudinal and circular polarizations will have
different masses, with the latter being dependent on the momentum.
In other words, instead of a "peak" corresponding to a given meson
we will see three "peaks" with the splitting depending on the value
of three-dimensional momentum. A similar phenomenon in a different
situation was obtained in Refs.~\cite{And2,And1}.

\section{Solvable models}

We have not yet specified the dilaton background $\varphi(z)$ and
the shape function $f(z)$ for the axial chemical potential.
Various holographic models for the "splitting phenomenon" can be
obtained by fixing these functions. The simplest solvable SW model
resulting in a Regge-like spectrum is given by the
background~\cite{son2}.
\begin{equation}
\label{11b}
\varphi=-\lambda^2z^2.
\end{equation}
Let us accept this background and consider the longitudinal
polarization. Following Ref.~\cite{son2}, we use the substitution
\begin{equation}
\label{12}
v(z)=e^{\lambda^2z^2/2}\sqrt{z}\psi(z),
\end{equation}
to convert the Eq.~\eqref{10} into the Schr\"{o}dinger form
\begin{equation}
\label{12b}
-\partial_z^2\psi_n^{\parallel}+\left(\lambda^4z^2+\frac{3}{4z^2}\right)\psi_n^{\parallel}=m_n^2\psi_n^{\parallel}.
\end{equation}
Introducing the dimensionless variable $y=\lambda z$,
Eq.~\eqref{12b} for the discrete mass spectrum $m_n^2=p_n^2$ of
the longitudinal (and time like) polarization transforms into
\begin{equation}
\label{13}
-\partial_y^2\psi_n^{\parallel}+\left(y^2+\frac{3}{4y^2}\right)\psi_n^{\parallel}=\frac{m_{n,\parallel}^2}{\lambda^2}\psi_n^{\parallel}.
\end{equation}
The normalized solutions are ($n=0,1,2,\dots$)
\begin{equation}
\label{14}
\psi_n^{\parallel}=\sqrt{\frac{2n!}{(1+n)!}}e^{-y^2/2}y^{3/2}L_n^1(y^2),
\end{equation}
where $L_n^1$ denote the associated Laguerre polynomials. The mass
spectrum is given by the corresponding eigenvalues,
\begin{equation}
\label{15}
m_{n,\parallel}^2=4\lambda^2(n+1),
\end{equation}
where the parameter $\lambda$ controls the slope of the radial
Regge trajectory.

To obtain the spectrum of circular polarizations we need to
specify the function $f(z)$. The choice of $f(z)$ must comply with
the requirement of correct UV/IR behavior of the action with
respect to conformal symmetry. This constraint leaves, however,
much freedom. We will be interested in the exactly solvable cases.
A simple possibility of this sort is given by the ansatz
\begin{equation}
\label{15b} f=\frac{b}{2}\frac{e^{-\lambda^2z^2}}{z}.
\end{equation}
Here $b$ is a dimensionless constant. After the
substitution~\eqref{12} the Schr\"{o}dinger equation~\eqref{13}
acquires an additional contribution,
\begin{equation}
\label{16}
-\partial_y^2\psi_n^{\pm}+\left(y^2+\frac{3}{4y^2}\pm\frac{b\mu_5}{\lambda^2}|\vec{p}|\right)\psi_n^{\pm}=
\frac{m_{n,\pm}^2}{\lambda^2}\psi_n^{\pm}.
\end{equation}
The eigenfunctions remain the same as in~\eqref{14} but the mass
spectrum is shifted,
\begin{equation}
\label{17}
m_{n,\pm}^2=4\lambda^2(n+1)\pm b\mu_5|\vec{p}|.
\end{equation}
Thus, the massive vector fields split into three polarizations
with masses $m_{n,-}<m_{n,\parallel}<m_{n,+}$. The mass splitting
is linearly dependent on the value of the spatial momentum
$\vec{p}$. The formula~\eqref{17} can be considered as a
generalization of the result of Ref.~\cite{And2} to the radially
excited spectrum.

The matching of~\eqref{17} with the corresponding expression in
Ref.~\cite{And2},
\begin{equation}
\label{18} m_{V,\pm}^2=m_V^2\pm \zeta|\vec{p}|,
\end{equation}
allows to estimate the constant $b$. Within the effective model
studied in Ref.~\cite{And2}, $\zeta=N_c g_\rho^2\mu_5/8\pi^2$,
where $g_\rho$ is related with the mass of vector particle,
$m_\rho^2=2g_\rho^2f_\pi^2\simeq m_\omega^2$. Substituting the
phenomenological values for $m_\rho$ and for the weak pion decay
constant $f_\pi$, one obtains the estimate $\zeta\simeq1.5\mu_5$.
We remark that this estimate can be made without phenomenological
values if one uses the formula for $m_\rho$ from the QCD sum rules
in the large-$N_c$ limit (see, e.g.,~\cite{sr}),
$m_\rho^2=24\pi^2f_\pi^2/N_c$ that yields $g_\rho=12\pi^2/N_c$ and
leads directly to $\zeta=\frac32\mu_5$. Comparing~\eqref{17}
and~\eqref{18} we arrive at the value $b=\frac32$.

Using the results of Ref.~\cite{gsw}, the background~\eqref{11b}
can be generalized to
\begin{equation}
\label{18b}
\varphi=-\lambda^2z^2\log U^2(c,0;\lambda^2z^2),
\end{equation}
where $U$ is the Tricomi hypergeometric function. Then the
dimensionless parameter $c$ in~\eqref{18b} will control the shift
of the radial Regge trajectories: $n+1$ in~\eqref{15}
and~\eqref{17} will be replaced by $n+1+c$.

The model was designed such that the relation
$m_{n,+}^2-m_{n,-}^2=\text{const}$ holds for any $n$. Other
possibilities are of course possible. For instance, one can
construct a solvable model with the relation
$m_{n,+}^2/m_{n,-}^2=\text{const}$. This is achieved via
replacing~\eqref{15b} by
\begin{equation}
\label{19}
f=\frac{\tilde{b}}{2}z e^{-\lambda^2z^2}.
\end{equation}
Introducing the variable
\begin{equation}
\label{19b}
y=\left(1\pm\frac{\tilde{b}\mu_5}{\lambda^4}|\vec{p}|\right)^{\frac14}\lambda z,
\end{equation}
the analogue of Eq.~\eqref{16} looks as follows
\begin{equation}
\label{20}
-\partial_y^2\psi_n^{\pm}+\left(y^2+\frac{3}{4y^2}\right)\psi_n^{\pm}=
\frac{m_{n,\pm}^2}{\lambda^2\sqrt{1\pm\frac{\tilde{b}\mu_5}{\lambda^4}|\vec{p}|}}\psi_n^{\pm},
\end{equation}
resulting in the mass spectrum
\begin{equation}
\label{21}
m_{n,\pm}^2=4\lambda^2\sqrt{1\pm\frac{\tilde{b}\mu_5}{\lambda^4}|\vec{p}|}\,(n+1).
\end{equation}
The eigenfunctions~\eqref{14} are now momentum- and
$\mu_5$-dependent because the variable $y$~\eqref{19b} depends on
$|\vec{p}|$ and $\mu_5$. The matching to Eq.~\eqref{18} for $n=0$
for small $|\vec{p}|$ gives $\tilde{b}=\frac34\lambda^2$. In this
scenario, the contribution of $\mu_5$ to the masses grows with
$n$.

Since the effective masses of polarized modes depend on the
spatial momentum in the models under consideration, a certain care
must be exercised to provide stability and, as a consequence,
to avoid the superluminal propagation. From~\eqref{17} we have for
the energy
\begin{equation}
\label{21b}
p_{0,n}^2=\vec{p}{\,}^2+4\lambda^2(n+1)\pm b\mu_5|\vec{p}|.
\end{equation}
In this respect, the $\varepsilon_-$ polarization is dangerous.
Imposing the condition $p_{0,n}^2\geq0$ we arrive at
\begin{equation}
\label{21c}
\mu_5^2\leq\frac{16\lambda^2}{b}(n+1).
\end{equation}
The strongest limitation comes from the ground state $n=0$,
\begin{equation}
\label{21d}
\mu_5\leq\frac{4\lambda}{b}.
\end{equation}

Consider now the group velocity $v=\frac{dp_0}{d|\vec{p}|}$. It
cannot exceed the speed of light in the vacuum, $v\leq1$.
For~\eqref{21b} this yields
\begin{equation}
\label{21e}
\frac{2|\vec{p}|\pm b\mu_5}{2\sqrt{\vec{p}{\,}^2+4\lambda^2(n+1)\pm b\mu_5|\vec{p}|}}\leq1.
\end{equation}
The constraint~\eqref{21e} leads to the same
limitations~\eqref{21c} and~\eqref{21d}.

The analogue of constraint~\eqref{21e} for the second model
considered above takes the form
\begin{equation}
\label{21g}
\frac{|\vec{p}|\pm \frac{b\mu_5(n+1)}{\lambda^2\sqrt{1\pm\,\tilde{b}\mu_5|\vec{p}|/\lambda^4}}}
{\sqrt{\vec{p}{\,}^2+4\lambda^2\sqrt{1\pm\tilde{b}\mu_5|\vec{p}|/\lambda^4}(n+1)}}\leq1.
\end{equation}
It is seen that the condition~\eqref{21g} cannot be fulfilled at
any $n$ --- the larger is $n$ the stronger is limitation on
$\mu_5$. Since $n$ labels the Kaluza--Klein modes of a
five-dimensional field, all infinite number of resonances must be
present simultaneously in agreement with the limit
$N_c\rightarrow\infty$. Thus, the second solvable model should be
disregarded.

\section{Concluding discussions}

According to some estimates of Ref.~\cite{And2}, the natural
value of $\mu_5$ in the central heavy ion collisions is in the few
hundreds of MeV. Such values are far below the
limitation~\eqref{21d},
\begin{equation}
\mu_5\leq
\frac{4\lambda}{b}\approx\frac{8\lambda}{3}\approx1400\,\text{MeV},
\end{equation}
where the value $\lambda\approx530$~MeV is taken from the typical
slopes of the vector radial trajectories. Thus, the first model
passes the stability criterium. We have shown that this criterium
imposes a strong constraint on the possible choice of models. In
particular, it seems to falsify the second model constructed in
the previous Section.

In principle, the effect of mass splitting between different
polarizations can be quite large. The masses of $\varepsilon_+$
and $\varepsilon_-$ circular polarizations of neighboring
resonances can even overlap.

We expect that the contribution of the axial chemical potential to
the hadron masses is approximately constant, i.e. it does not
depend on the radial number $n$. In this regard, the first
constructed model looks physical.

The splitting of effective masses of two polarizations signifies
the parity breaking in the medium. In the framework of
Ref.~\cite{And2}, this effect is related with the violation of 4D
Lorentz invariance due to the time-dependent background which is
present in the Lagrangian from the very beginning. In the
presented holographic description, we have essentially the same
situation.

The experimental implications of the presence of the axial
chemical potential are discussed in detail in Ref.~\cite{And3}.

Finally we mention some possible future directions. First of all,
it would be interesting to study the impact of the axial chemical
potential on the masses of vector mesons in the top-down
holographic approach. Second, it makes sense to include a non-zero
isospin chemical potential $\mu_I$ into our consideration. In the
low-energy QCD, $\mu_I$ contributes to the masses of charged
pions, $m_{\pi^\pm}\approx m_{\pi^0}\pm\mu_I$, and triggers the
pion condensation at values $\mu_I>m_{\pi^0}$~\cite{Son}. This
result is reproduced by the hard-wall bottom-up holographic
model~\cite{Albrecht} and by the Sakai-Sugimoto top-down
holographic model~\cite{Aharony}. In addition, the latter analysis
shows that at $\mu_I\gtrsim 1.7m_\rho$, the lowest vector meson
also condenses. It is thus interesting to verify this effect in
the bottom-up approach and then to study the interplay of $\mu_I$
and the axial chemical potential.

\section*{Acknowledgments}

The work was partially supported by the RFBR Grant No.
13-02-00127-a. S.S. Afonin acknowledges Saint-Petersburg State
University for Research Grant No. 11.38.189.2014 and Travel
Grant No. 11.42.1294.2014. D. Espriu
acknowledges the financial support from projects FPA2013-46570,
2014-SGR-104, CPAN (Consolider CSD2007-00042). D. Espriu would
like to thank the hospitality extended to him at the
University of Saint Petersburg. S.S. Afonin and A.A. Andrianov are
grateful for the hospitality extended to them at the
University of Barcelona.

\end{document}